# Edited Volumes, Monographs and Book Chapters in the Book Citation Index (BKCI) and Science Citation Index (SCI, SoSCI, A&HCI)

Loet Leydesdorff[i] & Ulrike Felt[ii]

**ABSTRACT**

In 2011, Thomson-Reuters introduced the Book Citation Index (BKCI) as part of the Science Citation Index (SCI). The interface of the Web of Science version 5 enables users to search for both "Books" and "Book Chapters" as new categories. Books and book chapters, however, were always among the cited references, and book chapters have been included in the database since 2005. We explore the two categories with both BKCI and SCI, and in the sister social sciences (SoSCI) and the arts & humanities (A&HCI) databases. Book chapters in edited volumes can be highly cited. Books contain many citing references but are relatively less cited. This may find its origin in the slower circulation of books than of journal articles. It is possible to distinguish between monographs and edited volumes among the "Books" scientometrically. Monographs may be underrated in terms of citation impact or overrated using publication performance indicators because individual chapters are counted as contributions separately in terms of articles, reviews, and/or book chapters.

[i] Amsterdam School of Communication Research (ASCoR), University of Amsterdam, Kloveniersburgwal 48, 1012 CX Amsterdam, The Netherlands; loet@leydesdorff.net ; http://www.leydesdorff.net
[ii] Department of Social Studies of Science, University of Vienna, Universitätsstraße 7/II/6 (NIG), 1010 Vienna, Austria; ulrike.felt@univie.ac.at

# INTRODUCTION

On the occasion of the inaugural issue of the *Journal of Scientometric Research*, let us turn to "Books" and "Book Chapters" as two new document types in the Web of Science (WoS). It has been argued that books and edited volumes are particularly important in the assessment of productivity and impact in the social sciences and humanities (e.g., Hammarfelt, 2011, 2012; Hicks, 2004; Larivière *et al.*, 2006; Leydesdorff *et al.*, 2010; Lindholm-Romantschuk *et al.*, 1996; Nederhof, 2006). As is well-known, the citation databases—Scopus and the Web of Science (WoS)—are based on scanning the journal literature for citations (Garfield, 1972), and thus the social sciences and humanities (SSH) are probably underrepresented in this literature (Kousha *et al.*, 2011; Kousha and Thelwall, 2009).

The new document types of "Books" and "Book Chapters" were made available as searchable fields with the introduction of version 5 of WoS in August 2011. In the second half of 2011, Thomson-Reuters (TR)—the present owner of the Science Citation Index (SCI)—also announced the introduction of a Book Citation Index (BKCI) as a complement to SCI (Adams & Testa, 2011). The BKCI would be launched with initial coverage of scholarly books published during the last 5 years in the Science edition and during the last 7 years in the editions for SSH. At the time of this research (March/April, 2012), the BKCI was not yet available at Dutch universities but we noted that the University of Vienna has already subscribed to BKCI.

On April 1, 2012, the Science edition of BKCI added 5,874 books and 179,906 book chapters to the SCI-Expanded edition and 12,706 books and 232,577 book chapters to the SSH edition. Each chapter of a book is processed separately in case of both edited volumes and monographs. One can sort the two types of books separately because all chapters in monographs most often have the same author name. (This criterion may not work when a colleague has written an editorial preface!) Book chapters are usually also attributed with another document type such as "article" or "review." These so-called document types are important for assessment since one has to control for document types (Garfield, 1979; Moed *et al.*, 1995; Schubert & Braun, 1986) in the evaluation. "Letters to the Editor," for example, are cited much faster than "Reviews" (Leydesdorff, 2008).

Before the introduction of version 5 of the WoS, one could already search the citations to book titles among the so-called "non-source literature references." Authors of articles included in the database could cite from all sorts of materials including books, patents and newspapers (Bensman & Leydesdorff, 2009; Nederhof *et al.*, 2010). BKCI, however, includes references within books in the source materials of the indices. Furthermore, it is seamlessly integrated at the WoS interface. As noted, availability depends on the institutional subscription. Results at one installation may therefore seem not reproducible at another installation of WoS.

Our project was triggered when one of us found 28 documents for an author when searching in Amsterdam and the other found 48 documents in Vienna. In addition to the 28 documents retrieved from the journal citation indices, the social scientist in question had authored a monograph and edited a book since 2005. However, we noted that "Books" and "Book Chapters" were also included in WoS as document types before the extension to BKCI. In this inaugural issue of the journal, let us explore these two document types in greater detail: How and since when have they been included in the citation indices at WoS? What are their scientometric charactistics?

**METHODS AND MATERIALS**

Searching the SCI-Expanded, SoSCI and the Arts & Humanities Citation Index (A&HCI), but without BKCI, provided us with a recall of 26 book titles and 19,017 book chapters on March 31, 2012. We used the search string "au = (a* or b* or c* or d* or e* of f* or g* or h* or i* or j* or k* or l* or m* or n* or o* or p* or q* or r* or s* or t* or u* or v* or w* or x* or y* or z*)." This search string would not retrieve documents without identifiable authorship. The two sets for books and book chapters do not overlap because they add up exactly when using an OR-statement.

One of us downloaded these materials from the WoS installation at the University of Amsterdam. In the following research we use various tools available for analyzing materials from WoS, for example, from http://www.leydesdorff.net/indicators or the analytical tools available at WoS for exploring the contents of these recalls.

**RESULTS**

*a. Time series*

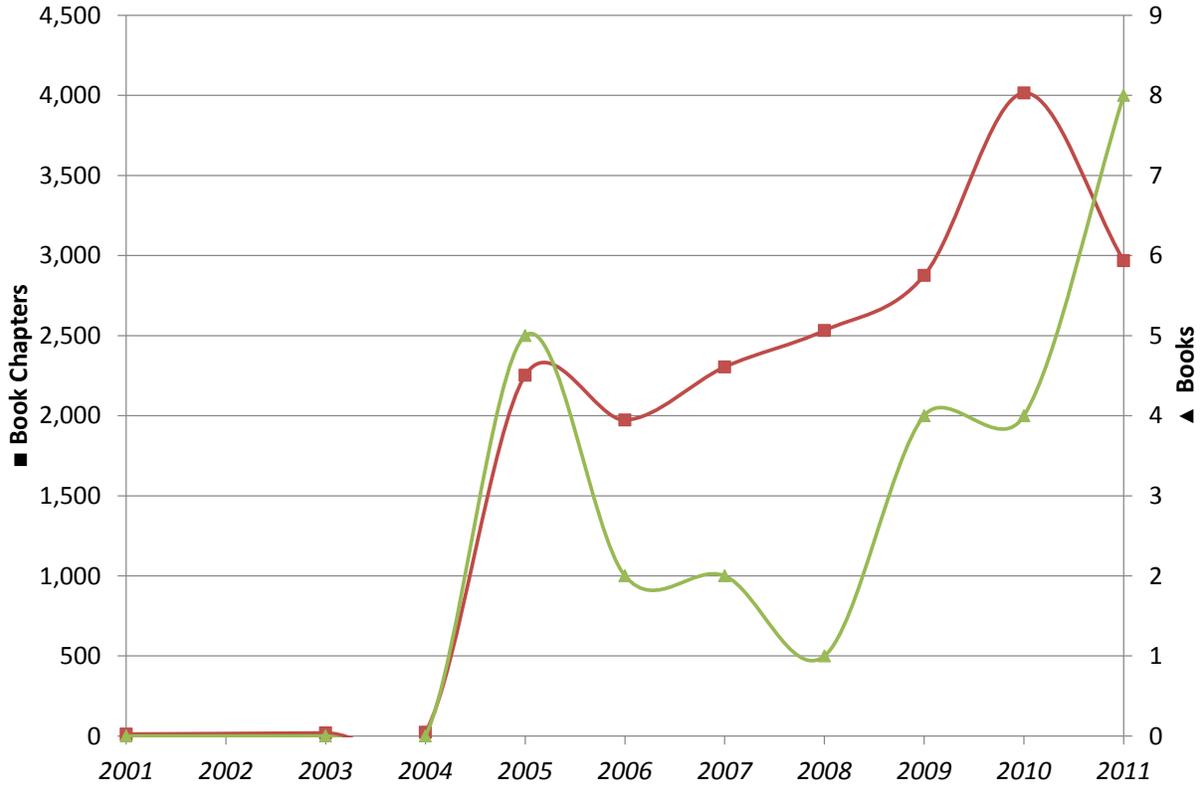

**Figure 1.** Books (▲) and Book Chapters (■) as Data Types in the SCI-E, SoSCI and A&HCI combined.

Figure 1 shows the time series of the two newly added document types in SCI-E, SoSCI and A&HCI without taking BKCI into account. "Books" did not occur at all as a document type before 2005 (right vertical axis); "book chapters" were only on the order of 10-20 before 2005, but since then, the recall became more than 2,000 (left vertical axis). The trends of both curves are somewhat upwards, but there are important irregularities in the years. In summary, the database is relevant from the perspective of our research question only since 2005. As noted,

BKCI—to be discussed below—includes "Books" and "Book chapters" only since 2005. Note that in the case of SSH, older literature may be as relevant as the most recent books.

*b. Books*

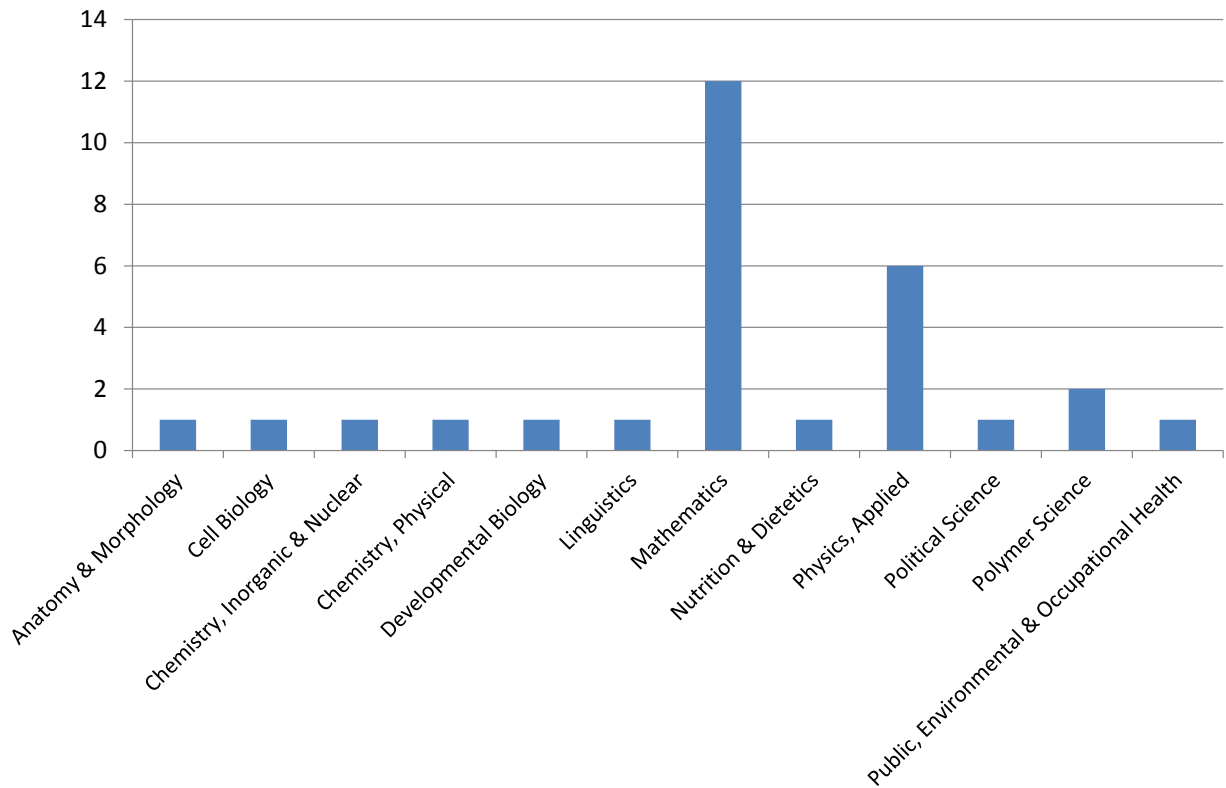

**Figure 2**: Disciplinary distribution of the 26 books retrieved in terms of the WoS Subject Categories.

Figure 2 provides the distribution of the 26 books retrieved, classified in terms of the WoS Subject Categories. (The WoS Subject Categories are renamed ISI Subject Categories in WoS version 4.) The main participation is from mathematics. Fifteen of these 26 books are also classified as articles; 10 as reviews and a single book entitled *Annual Review of Political Science* is classified uniquely as a book. Among the 302 documents that can be retrieved using *Annual Review of Political Science* as a journal name, 127 are classified as "Book chapters" but this single one is classified as a true "Book." If one looks it up, it is volume 12 of this annual review containing 28 articles (which can also be retrieved separately). In sum, this seems like an error in the database. "Books" were not a significant classifier before the addition of BKCI to WoS.

*c. Book chapters*

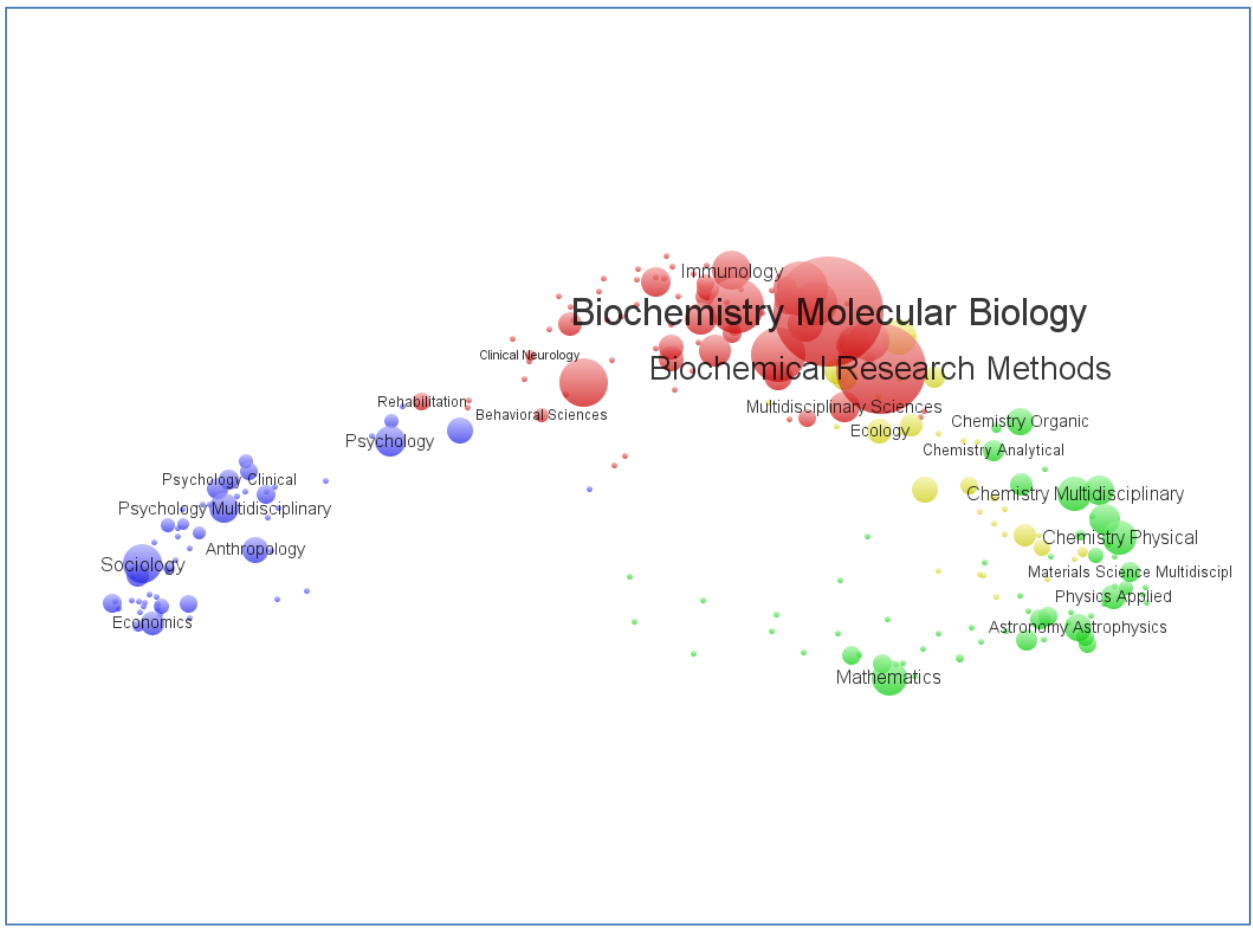

**Figure 3**: 19,017 book chapters in the SCI-Expanded, SoSCI and AHCI combined using the overlay map in VOSViewer (Leydesdorff *et al.*, under review).

In Figure 3, we used the 27,589 attributions of WoS SCs (by Thomson-Reuters) to the 19,017 "Book Chapters" retrieved and generated an overlay map using VOSViewer (Leydesdorff *et al.*, under review; Rafols *et al.*, 2010; see at http://www.leydesdorff.net/overlaytoolkit). The figure shows that "Book chapters" are common in a number of disciplines, but play an important role in the life sciences. Mathematics and Sociology, however, are also indicated on the map. Note that the overlay-map technique does not include the A&HCI (Leydesdorff *et al.*, 2011).

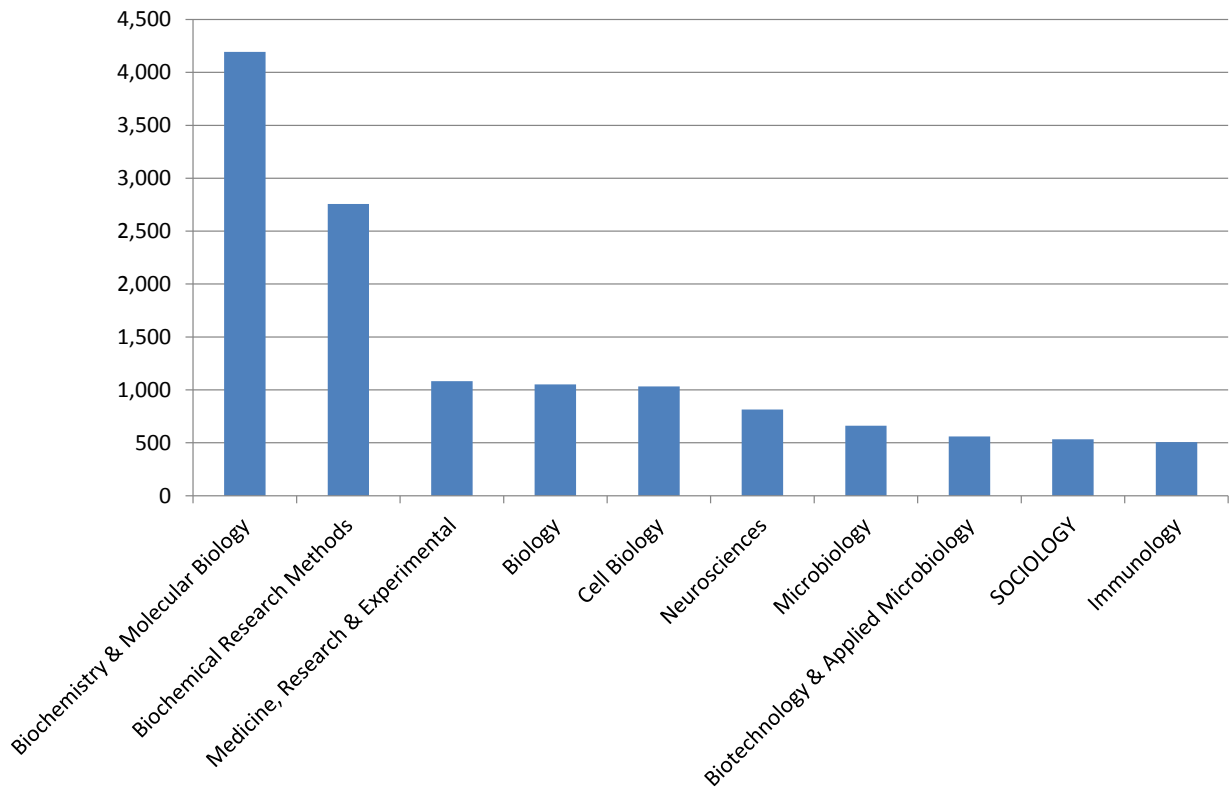

**Figure 4**: WoS Subject Categories with 500 or more book chapters among 19,017 book chapters retrieved from WoS.

Figure 4 shows the distribution for WoS SCs that were attributed 500 or more times to "Book chapters." The dominance of the biomedical sciences is clearly visible, but with 533 "Book chapters," "Sociology" is the leading non-biomedical field of science represented. The argument for the relevance of "Books" and "Book chapters" in analyzing and evaluating SSH is thus profiled. "Psychology" is the second largest group in SSH with 336 chapters.[1] "Linguistics," which can be considered as the most formalized discipline among the humanities (Leydesdorff *et al.*, 2011) follows with only 17 chapters.

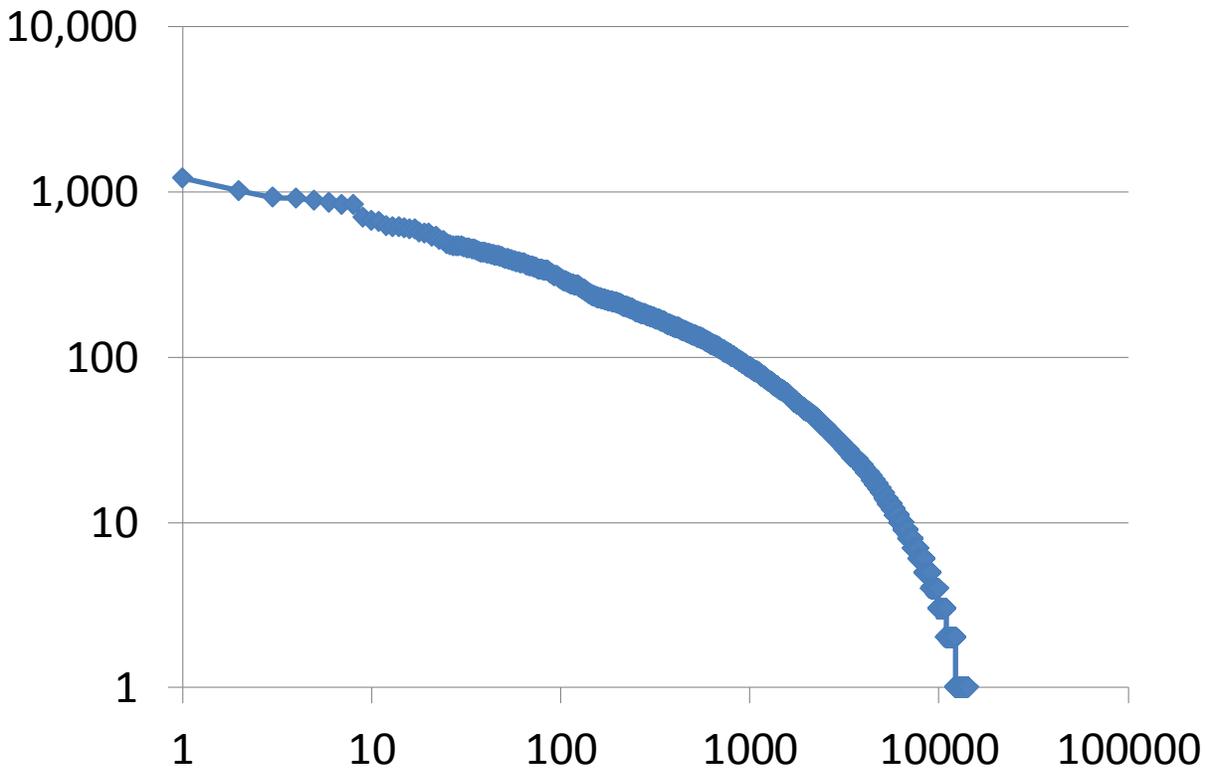

**Figure 5**: The citation distribution of 19,017 book chapters on a log-log scale.

The citation distribution of the book chapters is shown in a log-log format in Figure 5; 838 "Book chapters"—of which 817 are also classified by TR as reviews—are cited a hundred or more times.[2] This high citation score may partly be an effect of the specific distribution over the

---

disciplines. For example, citation rates are high among the bio-medical sciences (Garfield, 1979). However, "Book chapters" seem to be cited far more than an average article; the curve is not so skewed in the upper region, and only 5,807 "Book chapters" (30.5%) were never cited in these databases (SCI-Expanded, SoSCI and A&HCI, on March 31, 2012).

In summary, "Book chapters" (in edited volumes) have been an important medium of communication in a number of disciplines. Hitherto, they were not included separately among the source items for the computation of the impact factor, etc.; but 18,667 (98.2%) of the 19,017 indications of "Book chapters" were additionally indicated as "review" or "article." The others are also "editorial materials" or any of the remaining categories. "Book chapters" are rarely indicated without an additional attribution to another document type.

Let us finalize this analysis of the presence of "Books" and "Book chapters" in WoS before the introduction of BKCI with the following Table 1 which provides descriptive statistics for the parameters commonly used in scientometric analyses.

|  | **Books** | $N$ / **record** | **Book Chapters** | $N$ / **record** |
|---|---|---|---|---|
|  | (a) | (b) | (c) | (d) |
| $N$ of records | 26 |  | 19,016 |  |
| Authors | 73 | *2.8* | 48,785 | *2.6* |
| Institutional addresses | 36 | *1.4* | 34,261 | *1.8* |
| Cited references | 5,257 | *202.2* | 1,963,037 | *103.2* |
| Times cited (3/31/2012) | 272 | *10.5* | 366,826 | *19.3* |

**Table 1**: Descriptive statistics of scientometric parameters for "Books" and "Book Chapters" included in the SCI-E, SoSCI and AHCI combined.

In Table 1, the number of co-authors and institutional addresses per document is within the range of scientometric expectation. The number of references per document is large for "Book chapters," but twice as high for "Books." "Book chapters," however, are on average cited almost twice as often as "Books." The highest citation rate of a "Book" was only 71 as against almost a thousand "Book chapters" which were cited a hundred or more times.

## EXTENSION TO THE BKCI

Let us use the newly available BKCI to investigate this last conjecture that "Books" are not so highly cited as is often assumed. Their coverage by the citation indices might in that case not give such a strong boost to the citation scores in SSH as one might think when using Google Scholar. (Google Scholar has included "Books" since its launch in 2004; cf. Kousha & Thelwall, 2004.)

To this end, we downloaded the 12,706 books in the SSH Edition of BKCI and the 5,847 books in the Science Edition on April 2, 2012, using the same search string of all possible authors as above. These two sets contained an additional 245,252 and 185,767 book chapters, respectively. Furthermore, both sets contained approximately an additional 7,000 books without identifiable authorship. These are edited books often contained in book series and containing book chapters. The volume is then attributed as a book to the editors; the chapters can be classified as "Articles; Book chapter" and the introduction as "Editorial material; Book chapter". For reasons of consistency and because of our focus on the attribution of credit in the research question, we analyze the subsets of books that can be identified in terms of authorship.

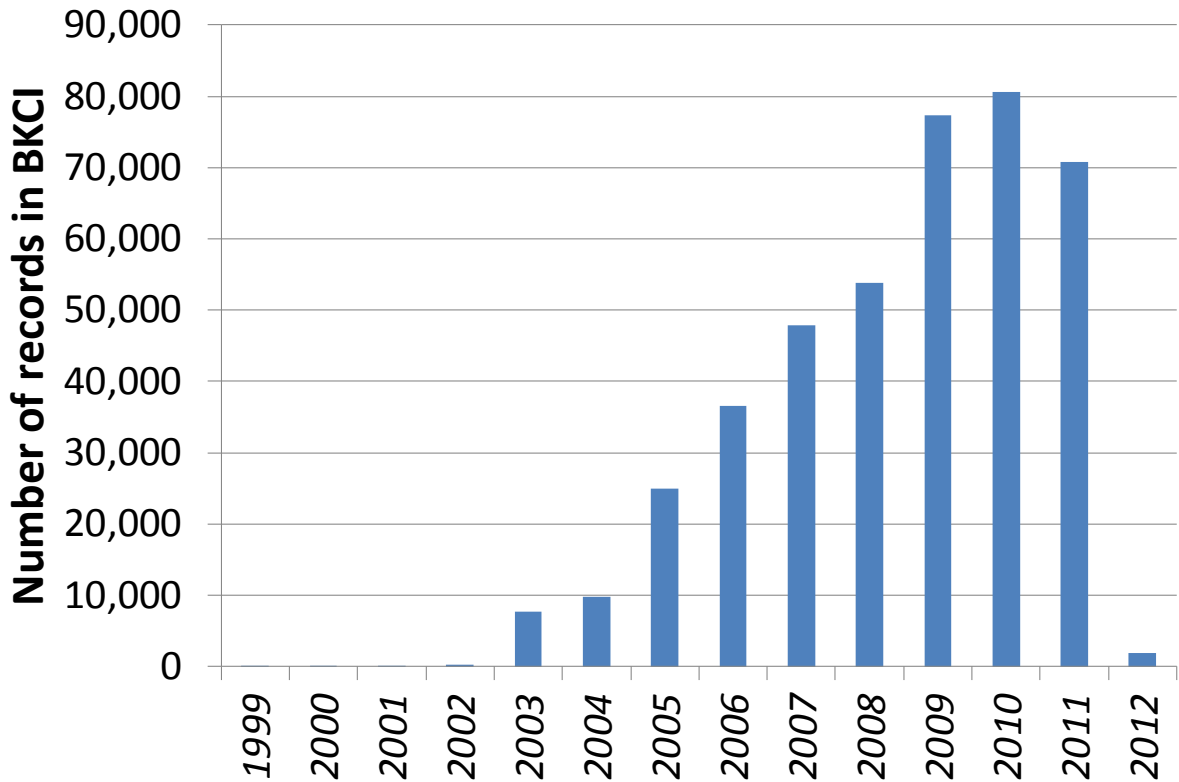

**Figure 6:** Distribution of all (411,712) records in terms of publication years in the BKCI (S and SSH combined).

Figure 6 first shows that there is a relevant filling of the database with "Books" and "Book chapters" having publication-year stamps previous to 2005, although the user interface at WoS indicates 2005 as the initial year. The difference between publication year and time of arrival at the office for data entry cannot explain this difference. Perhaps, one has to understand the message that BKCI is only reliable since 2005. The issue is relevant for a citation index since older publications have longer citation windows.

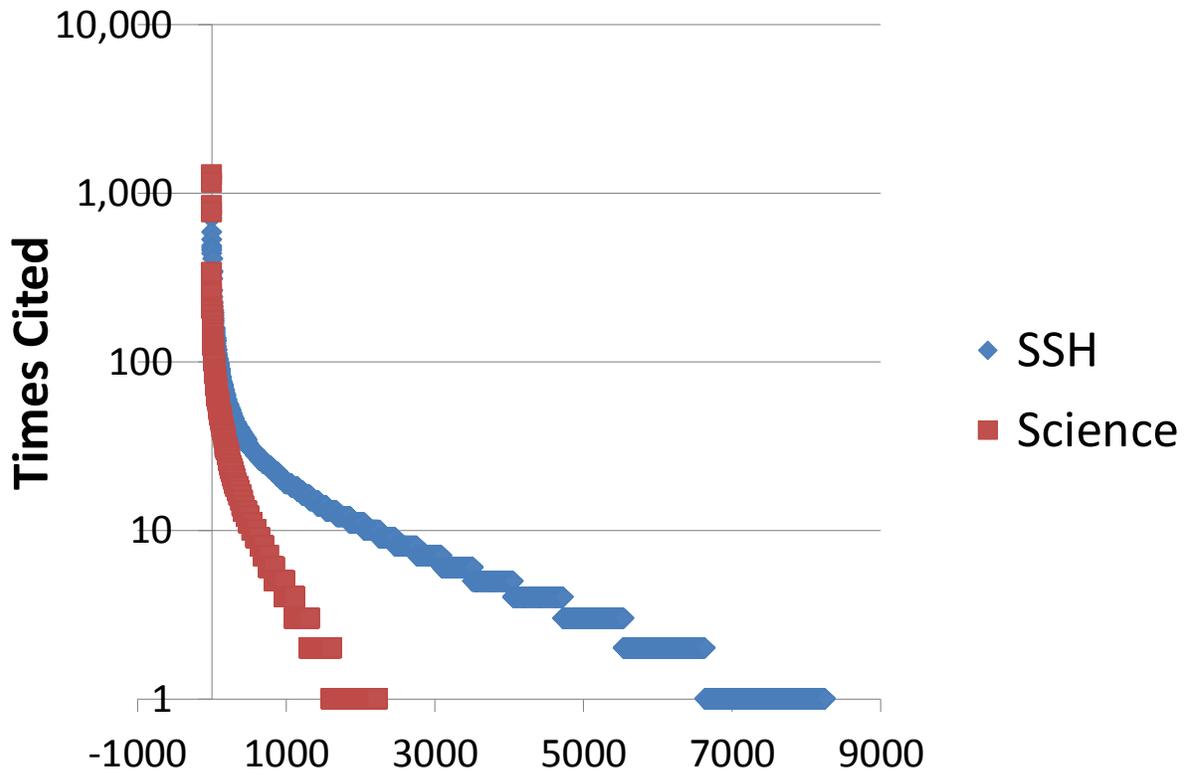

**Figure 7**: Times cited distributions of 12,706 and 5,874 books in the Book Citation Indices SSH and Science Edition, respectively (on April 2, 2012).

In Figure 7, we focus on the citation distributions of the two sets of books; 8,259 (65.0%) of the books in SSH were cited one or more times as compared to 2,238 (38.3%) in the Science Edition. However, these sets contain both monographs and edited volumes. In the case of edited volumes, one may prefer to cite chapters rather than the book title itself. More interesting is the percentage of books that are most highly cited; only 39 (0.7%) books in the Science Edition are cited a hundred or more times, whereas this number (percentage) is 91 (also 0.7%) for SSH.

| | Science Edition (a) | N / record (b) | SSH (c) | N / record (d) |
|---|---|---|---|---|
| N of records | 5,539[a] | | 12,706 | |
| Book Chapters | 185,767 | *31.6* | 245,252 | *19.3* |
| Authors | 9,813 | *1.8* | 15,777 | *1.2* |
| Institutional addresses | 957 | *0.2* | 557 | *0.0* |
| Cited references | 699,359 | *126.3* | 3,052,338 | *240.2* |
| Times cited (3/31/2012) | 29,233 | *5.3* | 89,774 | *7.1* |

[a] Only 5,539 of the 5,874 books were actually retrieved; the number of book chapters is normalized for 5,874, whereas the other parameters are normalized for 5,539.

**Table 2**: Descriptive statistics of scientometric parameters for Books in the Science and SSH Editions of the BKCI.

Table 2 provides the descriptive statistics. Proportionally more books are single-authored in the SSH set than in the Science Edition. Institutional addresses of authors are virtually absent. The number of references in SSH books is almost twice as high as in Science books. These figures accord with our intuitions but the relatively low citation rates of books came as a bit of a surprise.

We looked into the records of a single colleague (in the social sciences) who published two books in the period 2005-2011. The number of records for this author retrievable at WoS changes from 28 to 48 when the BKCI is included. As noted, the books are processed on a chapter-by-chapter basis. This accounts for 14 additional hits (five for one monograph and nine for an edited volume). The remaining six hits were due to including the database for conference proceedings which is available in Vienna but not in Amsterdam.

One of these two books is a monograph with four chapters, each of which is listed as a "Book chapter" while the monograph itself is listed as a "Book." In sum, this leads to five hits in the retrieval. None of the chapters contain any citing references (NRef = 0) while 111 references are integrated in a bibliography at the end of the book. The book itself was cited 13 times (since being published in 2008); none of the chapters was cited in this case. The author did not provide an institutional address inside the book, but on one of the flaps of the removable cover.

In the case of edited volumes, the same procedure is followed by Thomson-Reuters. The book is a single item and the eight chapters are included additionally. The book is cited nine times, but the chapters collect an additional 21 citations since publication in 2005. One of the chapters also obtained eight citations, while the citation rates for the others were below five. One could perhaps argue that a fraction of the credit for these chapters should be provided to the five editors of the volume. One of these co-editors is the author of the chapter that is cited eight times.

At Google Scholar, these two books were cited 16 and 21 times, respectively. The most highly cited chapter of the edited volume was now a different one cited 23 times, with 18 times for the runner-up. The co-editorship of the latter author, however, is only visible in two of the five so-called "related versions" at Google Scholar. This contribution is not visible using Publish-or-Perish as an interface to Google Scholar.[3] Using the same author as search identification in Scopus provides 30 documents but not the two books. As is well known, the journal coverage of Scopus is larger than WoS.

Thus, the issue of attributing citation credit to authors and/or editors can be confusingly complex. For example, Gorraiz & Grumpenberger (2012) noted most recently a monograph that received seven citations while the chapters had received 12 citations. One would need an additional routine to collect the citations to chapters in order to rank the books in the case of monographs.

---

[3] Publish-or-Perish is freeware for publication and citation analysis using Google Scholar, and is available at www.harzing.com.

**CONCLUSIONS AND DISCUSSION**

"Book chapters" can be considered as an additional categorization to a subset of articles and reviews. These chapters can be highly cited and contain on average a large number of references. "Books," however, could be considered as incidental classifications as before the introduction of the BKCI, they were not a relevant category.

The addition of the BKCI to WoS in 2011 has provided a seamless interface to WoS. When including book titles into the evaluation, one can distinguish between monographs and edited volumes in this database. However, it seems questionable that the credit for a monograph should depend on the organization of the book into chapters (given that each chapter counts as one publication). In other words, this may require normalization in addition to the control for document types and fields of science as is common in scientometric research (Leydesdorff *et al.*, 2011).

It might be useful to rethink the distinction between book series and annual series that are considered as part of the journal and series literature. The situation is further complicated because (book) series can contain both monographs and edited volumes. Anthologies would count as monographs when edited by the original author, but as edited volumes when edited by someone else (who may have added an introductory chapter that would count as "Editorial material" and hence be considered a non-citable item when an evaluator wishes to remain consistent with the definitions in use for the impact factor). In addition to repair work to the current edition of BKCI, one may wish to rethink the categories at the occasion of a next update (Jonathan Adams, *personal communication*, May 1, 2012).

Book citations are more scarce than one may have assumed. First, books circulate more slowly than journal literature. Reading books is time-consuming. This may particularly be a negative incentive in fields with research fronts and publication pressure such as biomedical sciences. In these fields, edited volumes are highly cited, but the indication "Book chapter" is additional to and thus covered by including reviews and articles as document types.

| Web of Science Categories | Record Count | % |
|---|---|---|
| Political Science | 31,112 | 7.55 |
| Economics | 24,684 | 5.99 |
| History | 22,499 | 5.46 |
| Education Educational Research | 20,426 | 4.96 |
| Biochemistry Molecular Biology | 13,986 | 3.39 |
| International Relations | 12,284 | 2.98 |
| Literary Theory Criticism | 12,173 | 2.95 |
| Business | 11,991 | 2.91 |
| Management | 11,665 | 2.83 |
| Philosophy | 11,328 | 2.75 |

**Table 3**. Top-10 WCs among the 412,039 "Books" and "Book Chapters" in BKCI (combined; 20 April 2012).

| Book | Times cited |
|------|-------------|
| Woodford, M, *Interest and Prices: Foundations of a Theory of Monetary Policy*, 2003. | 1244 |
| Nocedal, J, Wright, SJ, *Numerical Optimization*, Second Edition, 2006. | 775 |
| Gee, JP, *What Video Games Have to Teach Us About Learning and Literacy*, 2003. | 707 |
| Wegner, DM, *Illusion of Conscious Will*, 2002. | 590 |
| Slaughter, AM, *New World Order*, 2004. | 530 |
| North, DC, *Understanding the Process of Economic Change*, 2005. | 483 |
| Mesquita, BB, Smith, A, Siverson, RM, Morrow, JD, *Logic of Political Survival*, 2003. | 472 |
| Rose, N, *Politics of Life Itself: Biomedicine, Power, and Subjectivity in the Twenty-First Century*, 2007. | 460 |
| McNeil, AJ, Frey, R, Embrechts, P, *Quantitative Risk Management: Concepts, Techniques and Tools*, 2005. | 428 |
| Ostrom, E, *Understanding Institutional Diversity*, 2005. | 409 |

**Table 3**: Ten most highly cited books in BKCI-SSH on April 1, 2012.

Table 3 shows the predominance of the social sciences and the humanities in BKCI quantitatively. Table 4 lists the ten most highly cited books in the SSH across the disciplines and teaches us that among these, five are from years with publication dates older than 2005. As noted, the list in Table 4 should be understood as citations to full books; the possible citations to the chapters were not yet added in this case. Obviously, there is room for follow-up questions.

Let us finally note that BKCI does not include "citation classics" such as "Marx" or "Freud," since it reaches currently back to 2005 or in some case a few years more. These references are often important in SSH for intellectual reasons (Hammarfelt, 2011). Searching with "au = Marx K*" provides eight records of which seven are uncited translations of Marx' writings into English. An eighth record is a book chapter co-authored by Konstanze Marx (in German) and cited once.

**ACKNOWLEDGEMENT**

We are grateful to Juan Gorraiz, Jonathan Adams, and anonymous referees for relevant communications.